\documentclass{phb-proc4-auth}

\usepackage{graphicx}
\usepackage{amssymb}

\begin{document}
\begin{frontmatter}


\journal{SCES '04}


\title{Mott transitions in the multi-orbital systems}

%

\author[SW]{Akihisa Koga$^{a,}$\corauthref{1}}
\author[JPN]{Norio Kawakami}
\author[SW]{T.M. Rice}
\author[SW]{Manfred Sigrist}

 
\address[JPN]{Department of Applied Physics, 
Osaka University, Osaka 565-0871, Japan}
\address[SW]{Theoretische Physik, 
ETH-H\"onggerberg, 8093 Z\"urich, Switzerland}


\corauth[1]{Corresponding Author: Department of Applied Physics, 
Osaka University, Yamadaoka 2-1, Suita, Osaka, Japan 
Phone: +81-(6) 6879-7873 Fax: +81-(6) 6879-7873, Email: koga@tp.ap.eng.osaka-u.ac.jp}


\begin{abstract}
 We investigate the Mott transitions in the two-orbital Hubbard
model with different bandwidths. 
By combining dynamical mean field theory with 
the exact diagonalization,
we discuss  the stability of itinerant quasi-particle states in each band.
We demonstrate that separate Mott transitions occur at different
Coulomb interaction strengths in general, 
which merge to a single transition only under special conditions. 
In particular, it is clarified  that the $xy$ and pair-hopping
 components of the Hund coupling 
play a key role to control the nature of the Mott transitions.
\end{abstract}




\end{frontmatter}

%
%
%
%
%

Strongly correlated electron systems with multi-orbital bands pose a
variety of problems. 
One of the interesting problems is the orbital-selective Mott 
transition (OSMT) in the system with distinct orbitals.
The compounds $\rm Sr_2RuO_4$\cite{PT,RMP} and
$\rm La_{n+1}Ni_nO_{3n+1}$\cite{LaNiO} have the distinct type 
of orbitals in the $t_{2g}$ and $e_g$ bands, 
where the chemical substitution or the temperature should 
trigger an OSMT.
\cite{Nakatsuji,Kobayashi}
In contrast to these experimental findings, there still remains 
a theoretical controversy  for the nature of Mott transitions 
in the multi-orbital system.
Several groups claimed that the OSMT occurs 
in such a system. \cite{Anisimov,Fang,Koga,Sigrist}
On the other hand, Liebsch suggested
that a single Mott transition occurs in general. \cite{Liebsch}

A key to resolve the discrepancy may be the role
played by the Hund coupling $J$
 since the nature of the Mott transition crucially depends
on whether $J$ exists or not. \cite{Koga,Degenerate} As 
another remarkable point  on the Hund coupling, it is 
difficult to draw a definite conclusion on the ground-state
phase diagram by quantum Monte 
Carlo (QMC) simulations \cite{Liebsch} since the 
introduction of the Hund coupling
 (in particular its $xy$ 
and pair-hopping components)  causes serious sign 
problems at low temperatures.

In this paper, we revisit the OSMT in  the two-orbital 
Hubbard model with particular emphasis on 
the effect of the Hund coupling.  We clarify that 
the introduction of the anisotropy in the Hund coupling
 alters the nature of the Mott transitions.

Let us consider the following  two-orbital Hubbard Hamiltonian 
at half filling,
\begin{eqnarray}
H&=&\sum_{\stackrel{<i,j>}{\alpha,\sigma}}
\left(t_{ij}^{(\alpha)}-\mu\delta_{ij}\right)c_{i\alpha\sigma}^\dag
c_{j\alpha\sigma}
+U\sum_{i\alpha}n_{i\alpha\uparrow}n_{i\alpha\downarrow}\nonumber\\
&+&\left(U'-J\right)\sum_{i\sigma}n_{i1\sigma}n_{i2\sigma}
+U'\sum_{i\sigma}n_{i1\sigma}n_{i2\bar{\sigma}}\nonumber\\
&-&J_{xy}\sum_{i}\left[ c_{i1\uparrow}^\dag c_{i1\downarrow}
c_{i2\downarrow}^\dag c_{i2\uparrow}+c_{i1\downarrow}^\dag c_{i1\uparrow}
c_{i2\uparrow}^\dag c_{i2\downarrow}\right] \nonumber\\
&-&J'\sum_{i}\left[ c_{i1\uparrow}^\dag c_{i1\downarrow}^\dag
c_{i2\uparrow} c_{i2\downarrow}+ c_{i2\uparrow}^\dag c_{i2\downarrow}^\dag
c_{i1\uparrow} c_{i1\downarrow} \right]
\label{Hamilt}
\end{eqnarray}
where $c_{i\alpha\sigma}^\dag (c_{i\alpha\sigma})$ 
creates (annihilates) an electron 
with  spin $\sigma(=\uparrow, \downarrow)$ and orbital
index $\alpha(=1, 2)$ at the $i$th site and 
$n_{i\alpha\sigma}=c_{i\alpha\sigma}^\dag c_{i\alpha\sigma}$. 
$t_{ij}^{(\alpha)}$ denotes the hopping
integral for orbital $ \alpha $,
$\mu$ the chemical potential and
$U$ ($U'$) the intraband (interband) Coulomb interaction.

For the model with isotropic Hund coupling
($J=J_{xy}=J'$), it was shown  that the 
OSMTs occur for  $J \neq 0$ in general, which merge to 
a single transition only at $J=0$ under 
the condition $U=U'+2J$.\cite{Koga}  To clarify the
effect of the Hund coupling in more detail,
we  study the roles played by 
the $z$ component $J_z(=J)$, the $xy$ component $J_{xy}$, 
and the pair hopping $J'$.
In the following, we represent the two distinct electron bands by
semi-circular density of states (DOS),
$\rho_\alpha(x)=2/\pi D_\alpha \sqrt{1-(x/D_\alpha)^2}$,
where $2D_\alpha$ is the bandwidth.

\begin{figure}[htb]
\centering
\includegraphics[width=7cm]{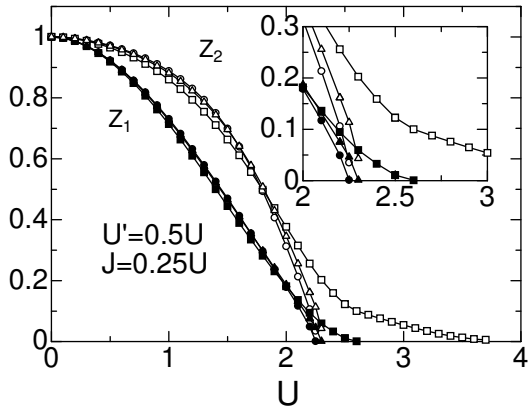}
\caption{Circles, triangles, and squares represent the 
quasi-particle weights 
for the case $J_{xy}/J=J'/J=0.0$, 0.5, and 1.0 
when $U'=0.5U, J=0.25U, D_1=1.0$ and $D_2=2.0$. 
} 
\label{fig:Z}
\end{figure}

By combining dynamical mean field theory\cite{Georges} 
with the exact diagonalization,\cite{Caffarel} 
we discuss the zero-temperature properties.
To clarify the stability of the paramagnetic metallic phase, 
we estimate the quasi-particle weight $Z_\alpha$ 
for each band ($\alpha=1,2$), as shown in Fig. \ref{fig:Z}. 
When $U=0$, the system is reduced to the two independent 
tight-binding models,
where $Z_\alpha=1$ for each orbital.
The introduction of the interactions
 results in the decrease of the weight for each band, 
for which the strong reduction  appears 
in the narrower band $(\alpha=1)$.  In this parameter region,
 the anisotropy of the Hund coupling only weakly
affects the behavior of the quasi-particle weights.
However, when the system approaches the critical point, 
the anisotropy drastically changes the values of $Z_{\alpha}$. 
When the system has the isotropic interactions $(J_{xy}=J'=J)$,
the Mott transitions occur at the different critical points 
depending on orbitals (i.e. OSMTs), as discussed in the 
previous paper. \cite{Koga}
On the other hand, away from the condition $(J_{xy}=J'=J)$,
the quasi-particle weights deviate dramatically
from the isotropic case,  
yielding very close transition points at $J_{xy}=J'=0$.
Therefore, we can say 
that the $xy$  as well as
 pair-hopping components of the Hund coupling 
are the relevant parameters to realize the OSMT.

In this connection, we wish to mention that the present 
results with $J_{xy}=J'=0$
exhibit quite similar properties found by Liebsch with QMC for
 the {\it isotropic} model with  $J=J_{xy}=J'$, 
based on which he concluded that the system  always
undergoes the single Mott transition.\cite{Liebsch} The present
analysis would suggest that in his analysis, the effect 
of  $J_{xy}=J'$ gets irrelevant by some technical reason
in QMC simulations, although
further detailed comparison should be necessary to draw the definite 
conclusion.

In summary we have  demonstrated that 
the  $xy$ and pair-hopping components of the Hund coupling 
are the key parameters to control the nature of the Mott transitions 
in the multi-orbital systems.
It is an interesting problem to explore the finite 
temperature properties 
around the critical point(s) in more detail, which is now under consideration. 


%
%
%
%

\end{document}